\documentclass[prl,twocolumn,superscriptaddress,showpacs]{revtex4}
\usepackage{bm}
\usepackage{amsmath}
\usepackage{amssymb}
\usepackage[dvips]{graphicx}
\begin{document}

\title{Aharonov-Bohm phase-driven resonant tunneling of interacting electrons in magnetopolaronic Majorana-Resonant-Level Model}

\author{Gleb A. Skorobagatko}

\email{gleb_skor@mail.ru}
\affiliation{ B.Verkin Institute for Low Temperature Physics \&
Engineering,  National Academy of Sciences of Ukraine, 47 Lenin
Ave, Kharkov, 61103, Ukraine}

\begin{abstract}
\begin{small}
The magnetopolaronic generalization of a Majorana-resonant-level (-MRL) model is considered for a single-level vibrating quantum dot symmetrically coupled to two half-infinite $g=1/2$- Tomonaga-Luttinger liquid (-TLL) leads at the Toulouse point. At the resonance by gate voltage the exact solution for the effective transmission coefficient is obtained in the whole range of magnetopolaronic coupling constant values. The obtained exact solution exists due to special Majorana-like symmetry of tunnel Hamiltonian and gives rise to nontrivial interference between different virtual vibronic channels of resonant tunneling with different fixed Aharonov-Bohm phases. This fact leads to a novel topologically nontrivial type of resonant Andreev-like magnetopolaronic tunneling in the system. As the result, in the zero-temperature limit, it is impossible to compensate the magnetopolaronic blockade in magnetopolaronic MRL-model by means of bias voltage, if vibron energy is the smallest (but nonzero) energy parameter in the system. 
\end{small}
\end{abstract}

\maketitle
\begin{normalsize}

Resonant tunneling in strongly interacting electron systems, particularly, in different types of molecular single-electron transistors (SET) still remains an attractive point in quantum mesoscopics \cite{PP,PT,NR,GR,GS,KF,FL,MCK,KG1,KG2,NEW,MY}. The molecular transistor in question is modelled as the single-level quantum dot (QD) vibrating along the $ 0y $ -axis in the transverse constant magnetic field \cite{OUR1,Pist}. Quantum dot is weakly coupled to two one-dimensional leads (quantum wires or carbon nanotubes) by means of two tunnel barriers \cite{KG1}. The half-infinite one-dimensional leads imply electron-electron interaction, which is described by Tomonaga-Luttinger liquid (-TLL) model with conventional TLL correlation parameter $g=(1+U_{TLL}/\pi v_{F})^{-1/2}$, ($0<g<1$) defined by the "bare" constant $ U_{TLL} $ of electron-electron interaction in TLL leads \cite{KR,AK}. For the most general situation of arbitrary $g$ and arbitrary magnetopolaronic coupling in quantum dot, it is impossible to solve the electron transport problem exactly \cite{KR}. In the most simple case of noninteracting (Fermi-liquid or FL-) leads (when $g=1$) both polaronic and magnetopolaronic SET models behave very similarly and it is difficult even to distinguish between them \cite{OUR1, Pist}. From the other hand, as it was shown earlier for the special value: $g=1/2$ of TLL correlation parameter, in the absence of any quantum vibrations of QD, the problem of resonant electron tunneling is exactly solvable even in the case of asymmetric tunnel coupling \cite{KF,KG1}. In that model, in the case of symmetric tunnel couplings, the Majorana-like symmetry emerges in the tunneling Hamiltonian and this is known as the TLL-realization of the Majorana-resonant-level model (MRLM) or simply as the spinless Tomonaga-Luttinger liquid resonant-level (TLLRL-) model \cite{AK}. Recently, it was shown by the author for the polaronic generalization of TLLRL- model (see \cite{MY}), that in the case of strong electromechanical coupling a novel type of Andreev-like resonant polaron-assisted tunneling is realized in the system. The latter polaronic $ g=1/2 $-TLLRL model strongly differs from the case of SET with noninteracting (FL-) leads \cite{MY,NEW}. Thus, one may ask if the magnetopolaronic TLLRL-model differs from polaronic one or, as it takes place in noninteracting case, these models are qualitatively similar? - Below the unexpected answer on this question will be given.

In this letter, it is shown for the first time for SET model with arbitrary magnetopolaronic coupling, that in the case of symmetric tunnel couplings between vibrating quantum dot and $g=1/2$- TLL leads, at the Toulouse point in Coulomb interaction strength (between TLL-leads and quantum dot) i.e. in the MRLM case \cite{AK,KG1} - the model become exactly solvable with respect to average current at the resonance by gate voltage. As the result, the exact formula for transmission coefficient of strongly correlated electrons is derived for magnetopolaronic MRL-model.

It is reasonable to start from the Hamiltonian of magnetopolaronic $ g=1/2 $-TLLRL model  already in its re-fermionized form (one can see corresponding unitary transformations in Refs.\cite{MY,KG1}):
\begin{equation} \label{1}
\hat{H}=\hat{H}_{l}+\hat{H}_{d}+\hat{H}_{t}.
\end{equation}
Here the first term describes the quadratic Hamiltonian of half-infinite one-dimensional $ g=1/2 $-TLL leads: $\hat{H}_{l}=\sum_{\pm}1/2\pi\int dx(\partial_{x}\Phi_{\pm}(x))^{2}$. (Here and below we put $\hbar v_{g}=\hbar v_{F}/g=1$ with "bare" Fermi velocity $v_{F}$.) The transformed bosonic phase fields: $\Phi_{\pm}(x)$ are connected with "initial" bosonic phase fields $\Phi_{j}(x)$ ($j=L,R$) of chiral (left- or right-moving) plasmonic charge density excitations in the $j$-th lead ($j=L,R$) by relation: $\Phi_{\pm}(x)=(\Phi_{L}(x)\pm\Phi_{R}(x))/\sqrt{2}$ (see Ref.\cite{KG1} for details). At points $x=\pm\infty$ the 1D half-infinite LL leads are coupled to two reservoirs of noninteracting electrons. The difference between chemical potentials of these reservoirs is proportional to bias voltage $ V $ being applied to the leads at points $x=\pm\infty$. Second term in Eq.(1): $\hat{H}_{d}=\Delta\hat{d}^{+}\hat{d}+\frac{\hbar\omega_{0}}{2}(\hat{p}_{y}^{2}+\hat{y}^{2})$ -represents the transformed Hamiltonian of single-level vibrating quantum dot (QD) at the Toulouse point in Coulomb interaction between QD and TLL leads (see Refs. \cite{KG1,KG2}). In the latter equality: $\hat{d}^{+}(\hat{d})$ - are the standard fermionic creation (-annihilation) operators on resonant level of QD; $\Delta=\Delta(V_{g})$ - is the resonant level energy, driven by $ V_{g} $-gate voltage being applied to QD electrostatically \cite{VR}.  In the case of MRL-coupling one should put: $ \Delta(V_{g})=0 $. One can always satisfy the latter "resonance" condition since $ V_{g} $ is an independent parameter in the model. Thus, in the magnetopolaronic MRLM case of interest one can write down  
\begin{equation} \label{2}
\hat{H}_{d}=\hat{H}_{v}=\frac{\hbar\omega_{0}}{2}(\hat{p}_{y}^{2}+\hat{y}^{2})
\end{equation}
i.e. the transformed Hamiltonian of fermionic resonant level of QD in the magnetopolaronic MRLM case contains only vibronic degrees of freedom. In Eq.(2) $\hbar\omega_{0}$ is the energy of vibrational quantum and $\hat{p}_{y}$,$\hat{y}$ are the dimensionless bosonic operators of the momentum and center-of-mass coordinate of QD in $ y $-direction. 
Then transformed tunnel Hamiltonian (the third term in Eq.(1)) takes the form:
\begin{eqnarray}\label{3}
\nonumber
\hat{H}_{t}=\hat{d}^{+}[\gamma_{L}\hat{X}_{L}^{+} \hat{\Psi}_{-}(0)+\gamma_{R}\hat{X}_{R}^{+} \hat{\Psi}_{-}^{+}(0)]+\nonumber\\
+[\gamma_{L}\hat{X}_{L} \hat{\Psi}_{-}^{+}(0)+\gamma_{R}\hat{X}_{R} \hat{\Psi}_{-}(0)]\hat{d},
\end{eqnarray}
where $\gamma_{L(R)}$ are the tunneling amplitudes for the left- and right TLL lead correspondingly. (In our notations: $\gamma_{L}^{2}+\gamma_{R}^{2}=\Gamma_{0}$ - the "bare" width of the fermionic level of QD in standard wide-band-limit (-WBL) approximation \cite{KG1}.) Operators $\hat{\Psi}_{\pm}(x)=\exp(i\Phi_{\pm}(x)/\sqrt{g})/\sqrt{2\pi a_{0}}$ stand for new fermions (being spatially nonlocal in $x$-direction) and fulfill standard fermionic anticommutation relations $\{\hat{\Psi}_{\pm}(x),\hat{\Psi}_{\pm}^{+}(x^{'})\}=\delta(x-x^{'})$ \cite{KG1}. The novel element in Eq.(3), as compared with conventional exactly solvable $ g=1/2 $-TLLRL model of Ref.\cite{KG1} without quantum vibrations, is the "magnetopolaronic" renormalization  \cite{OUR1} of tunneling amplitudes by bosonic operators:
\begin{equation} \label{4}
\left\{    \begin{array}{ll}
         \hat{X}_{L}=\exp(-i\phi\hat{y})  \\
          \hat{X}_{R}=\exp(i\phi\hat{y})
           \end{array}
          \right.
\end{equation}
 -which describe the influence of fluctuating Aharonov-Bohm phase, acquired by the electron in the process of resonant tunneling \cite{OUR1}. In Eq.(4) $ \phi=\Phi/\sqrt{2}\Phi_{0}=e y_{0}D_{0} H/\sqrt{2}hc$ - is the dimensionless magnetopolaronic coupling constant, where: $ y_{0}=\sqrt{\hbar/M\omega_{0}} $ -is the amplitude of zero-point oscillations of the QD center-of-mass coordinate in the  $ y $ -direction ($ M $ -is the mass of quantum dot); $ D_{0} $ - is the characteristic distance between two TLL leads (i.e. the characteristic size of QD region in the $ x $-direction); $ H $ - is the absolute value of constant external transverse magnetic field, which is nonzero only in the region of the length  $ D_{0} $ between two TLL electrodes. Obviously, bosonic operators in Eq.(4) have following symmetry: $ \hat{X}_{L(R)}^{+}=\hat{X}_{R(L)} $ and also
\begin{equation} \label{5}
\left( \hat{X}_{L(R)}\Leftrightarrow \hat{X}_{R(L)}\right) = \left( \hat{y}\Leftrightarrow -\hat{y} \right)
\end{equation}
-this symmetry, as it will be clear below, fixes definite values of relative Aharonov-Bohm phase of tunneling electron in each vibronic channel. 
Further, as it was mentioned in Ref. \cite{KG1}, since the "charge"-density $\Phi_{+}(x)$ - channel is decoupled at the Toulouse point, one can rewrite the current operator by means of only the "current"-density channel $\Phi_{-}(x)$ 
\begin{equation} \label{6}
 \hat{I}(\infty) =G_{0}[\hat{\Psi}_{-}^{+}\hat{\Psi}_{-}(-\infty)  -  \hat{\Psi}_{-}^{+}\hat{\Psi}_{-}(+\infty)],
\end{equation}
where $ G_{0}=e^{2}/h $-is the conductance quantum. 

Now to solve the model of Eqs.(1-6) a well-known quantum equation of motion (QEM) method could be used. The Heisenberg equations for fermionic operators take the form (at $ \Delta=0 $)
\begin{equation} \label{7}
i\hbar\partial_{t}\hat{d}=\gamma_{L}\hat{X}_{L}^{+} \hat{\Psi}_{-}(0)+\gamma_{R}\hat{X}_{R}^{+} \hat{\Psi}_{-}^{+}(0)
\end{equation}
\begin{equation} \label{8}
i\hbar\partial_{t}\hat{\Psi}_{-}(x)=-i\partial_{x}\hat{\Psi}_{-}(x)+\delta(x)[\gamma_{L}\hat{X}_{L}\hat{d}-\gamma_{R}\hat{X}_{R}^{+}\hat{d}^{+}],
\end{equation}
where, following Refs.\cite{KG1,KG2} we defined
$\hat{\Psi}_{-}(0)=(\hat{\Psi}_{-}(0^{-})+\hat{\Psi}_{-}(0^{+}))/2$, and $\delta(x)$ is the delta function. Integrating Eq.(8) in the small vicinity of the point $x=0$, one obtains
\begin{equation} \label{9}
i[\hat{\Psi}_{-}(0^{+})-\hat{\Psi}_{-}(0^{-})]=\gamma_{L}\hat{X}_{L}\hat{d}-\gamma_{R}\hat{X}_{R}^{+}\hat{d}^{+}.
\end{equation}
In the absence of magnetopolaronic coupling ($\hat{X}_{L(R)}^{+}=\hat{X}_{L(R)}=1$), Eqs.(7-9) are reduced to Eqs.(6) from Ref.\cite{KG1}. Integrating formally Eq.(7)(similarly to Ref.\cite{MY}) and substituting the obtained solution into Eq.(9) one can derive following basic integral operator equation in the form of the operator-valued boundary condition at physical point $ x=0 $: 
\begin{eqnarray}
   \hbar\lbrace\hat{\Psi}_{-}(0^{+};t)-\hat{\Psi}_{-}(0^{-};t)\rbrace=\nonumber \\
  - \lim_{\alpha \rightarrow 0}\int_{0}^{t}dt^{'}\{[\gamma_{L}^{2}\hat{X}_{L}(t)\hat{X}_{L}^{+}(t^{'})\hat{\Psi}_{-}(0;t^{'})
\nonumber \\
+\gamma_{L}\gamma_{R}\hat{X}_{L}(t)\hat{X}_{R}^{+}(t^{'})\hat{\Psi}_{-}^{+}(0;t^{'})]
e^{-\alpha(t-t^{'})/\hbar}\nonumber\\
+[\gamma_{R}^{2}\hat{X}_{R}^{+}(t)\hat{X}_{R}(t^{'})\hat{\Psi}_{-}(0;t^{'})\nonumber \\
 +\gamma_{L}\gamma_{R}\hat{X}_{L}^{+}(t)\hat{X}_{R}(t^{'})\hat{\Psi}_{-}^{+}(0;t^{'})] e^{-\alpha(t-t^{'})/\hbar}\}.
 \label{10}
\end{eqnarray}
Now, central operator equation (10) should be complemented by equation for bosonic operator $\hat{y}$. The corresponding Heisenberg equation could be rewritten in the form of the Newton-like equation of motion for operator $\hat{y}$ with quantum analog of Lorentz force in the right-hand side of the equation:
\begin{equation} \label{11}
[\partial_{t}^{2}+\omega_{0}^{2}]\hat{y}=-\phi\omega_{0}\hat{I}(0)=\phi\omega_{0}[\hat{\Psi}_{-}^{+}\hat{\Psi}_{-}(0^{+})  -  \hat{\Psi}_{-}^{+}\hat{\Psi}_{-}(0^{-})].
\end{equation}
Here $ \hat{I}(0) $ denotes the "MRL-current" operator at physical point $ x=0 $, i.e. at the boundary with quantum dot. One can check out, using Eq.(9) and canonical anticommutation relations for fermionic operators, that: $ i\hbar\partial_{t}\hat{I}(0)=-[\hat{\tilde{H}}_{t}, \hat{I}(0)] \propto -(\gamma_{L}^{2}- \gamma_{R}^{2})$. Consequently, in the MRLM case, where $\gamma_{L}=\gamma_{R}=\sqrt{\Gamma_{0}}/2$, one obtains: 
\begin{equation} \label{12}
i\hbar\partial_{t}\hat{I}(0)_{MRLM}=-[\hat{\tilde{H}}^{symm}_{t}, \hat{I}(0)]=0.
\end{equation}
Thus, in the magnetopolaronic MRLM case, the "quantum Lorentz force" operator does not depend on time, i.e. it does not fluctuate. Of course, the latter statement is not the case for $ \hat{I}(\infty) $ which represents "true" current operator in given realization of MRL-model \cite{CUR}. Since in symmetric case the right-hand side of Eq.(11) represents a constant operator, it could be excluded from Eq.(11) by the coordinate shift, which does not affect the resulting formulas. As the result, at $\gamma_{L}=\gamma_{R}$, Eq.(11) has a "free" solution of the form: $\hat{y}(t)=\frac{1}{\sqrt{2}}(\hat{b}_{0}^{+}e^{i\omega_{0}t}+\hat{b}_{0}e^{-i\omega_{0}t}) $  (bosonic operators $\hat{b}_{0}^{+}$($\hat{b}_{0}$) describe the creation (annihilation) of a free vibron and fulfill standard bosonic commutation relation $[\hat{b}_{0},\hat{b}_{0}^{+}]=1$). This solution  means that, in symmetric MRLM case, all averages with total transformed Hamiltonian (1) are factorized exactly on fermionic and bosonic parts. Thus, one can replace all the products of nonlinear bosonic operators $ \hat{X}_{L(R)}^{+}(t^{'})\hat{X}_{L(R)}(t) $ and $\hat{X}_{R(L)}^{+}(t^{'})\hat{X}_{L(R)}(t)$ in the basic operator equation (10) by corresponding averages with quadratic Hamiltonian of a decoupled quantum harmonic oscillator (2).
Following the method of Refs.\cite{KG1,MY}, which turns out to be exact in our case of magnetopolaronic MRL-model at resonance by gate voltage (at $ \Delta=0 $), one can rewrite the decomposition for $\hat{\Psi}_{-}(x,t)$ fermionic operators from Ref.\cite{KG1}: 
\begin{equation} \label{13}
\hat{\Psi}_{-}(x;t)=\int\frac{dk}{2\pi}e^{ik(t-x)}\left\{
                                                    \begin{array}{ll}
                                                      \hat{a}_{k}, & x<0  \\
                                                      \hat{b}_{k}, & x>0
                                                    \end{array}
                                                  \right.
\end{equation}
(where: $\hat{a}_{k}^{+}$ ($\hat{a}_{k}$) are the standard fermionic creation (annihilation) operators. (Note, that $ \hat{b}_{k}=t(k)\hat{a}_{k} $, where $ t(k) $ is the transmission amplitude.) After that, similarly to Refs. \cite{KG1,MY}, using Eqs.(6),(12-13) one can write down the Landauer-type transport formula for average current in the magnetopolaronic MRL-model: 
\begin{equation}\label{14}
\langle\hat{I}(\infty)\rangle=G_{0}\int d\varepsilon R_{\phi0}(\varepsilon)[n_{F}(\varepsilon-eV)-n_{F}(\varepsilon)],
\end{equation}
where $R_{\phi0}(\varepsilon)=1-|t(\varepsilon)|^{2}$ is the energy-dependent coefficient of "Andreev-like" reflection of $\hat{\Psi}_{-}$- fermions, which determines the transmission coefficient for physical electrons and $ n_{F}(\varepsilon)=(e^{\beta\varepsilon}+1)^{-1} $ is the Fermi-Dirac distribution function ($ \beta^{-1}=T $ is the temperature). Starting from here, it is reasonable to put everywhere: $\gamma_{L}=\gamma_{R}=\sqrt{\Gamma_{0}}/2$, \cite{PTA}. Then, solving the basic operator equation (10) together with its hermitian-conjugated equation and with Eq.(13), one can derive following \textit{exact} formula for the transmission coefficient of our magnetopolaronic MRL-model for arbitrary value of magnetopolaronic coupling constant $\phi$: 
\begin{equation} \label{15}
 R_{\phi0}(\varepsilon)=\frac{(2\tilde{\Gamma}_{\phi})^{2}}{1+2(\tilde{\Gamma}_{N}^{2}+\tilde{\Gamma}_{\phi}^{2})+(\tilde{\Gamma}_{N}^{2}-\tilde{\Gamma}_{\phi}^{2})^{2}}
\end{equation}
with
\begin{equation}\label{16}
\tilde{\Gamma}_{N}(\varepsilon)=\sum_{l=-\infty}^{+\infty}\frac{\Gamma_{0}F_{l}(\beta)\varepsilon}{(\hbar\omega_{0}l)^{2}-\varepsilon^{2}} ; 
\end{equation}
\begin{equation} \label{17}
\tilde{\Gamma}_{\phi}(\varepsilon)=\sum_{l=-\infty}^{+\infty}\frac{(-1)^{l}\Gamma_{0}F_{l}(\beta)\varepsilon}{(\hbar\omega_{0}l)^{2}-\varepsilon^{2}};
\end{equation}
Here $ F_{l}(\beta)=e^{-\phi^{2}(1+2n_{b})}I_{l}(\phi^{2}\sqrt{n_{b}(1+n_{b})})e^{-\beta \hbar\omega_{0} l/2} $ and $I_{l}(z)$ is the Bessel function of $l$-th order of the imaginary argument, $ n_{b}=(e^{\beta \hbar\omega_{0}}-1)^{-1} $ is the Bose-Einstein distribution function. Formulas (15-17) are \textit{exact} and represent the central result of this letter. Obviously, the most distinctive quantum features of the model manifest itself at low temperatures, when: $ T\ll \hbar\omega_{0}$. First, at $ \phi=0$ and $\hbar\omega_{0}=0 $ formulas (15-17) result in: $ R_{0}= \Gamma^{2}_{0}/(\varepsilon^{2}+\Gamma^{2}_{0})$ -a well-known Breit-Wigner transmission coefficient for resonant tunneling through $ g=1/2 $- TLLRL model \cite{KF,KG1}. In the case where: $ \Gamma_{0} \ll \hbar\omega_{0} \ll T $ (i.e. in the limiting case of "sequential" electron tunneling \cite{MY}), when: $ \phi^{2}\ll 1  $, from formulas (15-17) one obtains, that the height of the $ l $-th satellite peak of low-temperature differential conductance (at $ eV=\hbar\omega_{0} \vert l \vert $) is proportional to $ (\phi^{2})^{\vert l \vert} $, in according with standard predictions of perturbation theory in $ \phi^{2} $. Further, the most general realization of limiting case: $ \Gamma_{0}\ll \hbar\omega_{0} $ (and $\hbar\omega_{0} \gg T  $), for strong magnetopolaronic coupling ( $ \phi^{2} \geq 1  $) - is depicted on Fig.1,a (blue solid line), in comparison with corresponding polaronic MRLM case at electromechanical coupling of the same strength (red dotted line). One can see from this figure, that difference between polaronic and magnetopolaronic MRL-models is due to interference between different virtual channels of resonant tunneling in the magnetopolaronic MRLM case. This interference could be either constructive or destructive, depending on the energy of the incoming quasiparticle. Since such interference is absent in the "convinient" magnetopolaronic FLL-model of Ref.\cite{OUR1}, it is believed to be specific only for MRL-type of magnetopolaronic coupling. Other important limit (at zero temperature): $ \Gamma_{0}\gg \hbar\omega_{0} \rightarrow 0 $ ($ T< \hbar\omega_{0} $) of general formulas (15-17) turns out to be even more interesting. In this case all virtual vibronic channels give contribution to  infinite sums of Eqs.(16),(17). As the result, in the limit $ \hbar\omega_{0} \rightarrow 0 $ one can easily calculate sums (16),(17) and write down following explicit formula for the transmission coefficient:
\begin{equation}\label{18}
R^{(0)}_{\phi0}(\varepsilon)=\frac{(4\Gamma^{2}_{0}e^{-4\phi^{2}})\varepsilon^{2}}{\varepsilon^{4}+2\Gamma^{2}_{0}(1+e^{-4\phi^{2}})\varepsilon^{2}+\Gamma^{4}_{0}(1-e^{-4\phi^{2}})^{2}}.
\end{equation}  
Remarkably, in the limit: $ \phi^{2}\ll 1 $ the above formula (18) totally coincides (up to redefinition of $ \Gamma_{0} $ and $ \phi $) with formula (16) from Ref. \cite{MK} for effective transmission coefficient $ D^{eff}_{0}(\omega) $, which has been calculated for polaronic MRLM model in the limits of perturbation theory in small constant of electromechanical coupling by means of the full-counting statistics (FCS) method. However in the magnetopolaronic MRLM case, Eq.(18) remains valid in a whole range of $ \phi^{2} $ values. On Fig.1,b, the transmission coefficient (18) is plotted for two important cases: i) when $ \phi=0$ (red solid line) and ii) when $ \phi\neq 0$, ($ \phi=0.5$ -blue solid line). From Fig.1,b one can see the sharp difference between these two cases. Indeed, in the case $ \phi=0$ we have a usual Lorenzian-shape curve for transmission coefficient, while in the case $ \phi\neq 0$ (even at $ \phi\ll 1 $) the interference results in a strongly nonmonotonic behaviour of transmission coefficient. As the result, transmission coefficient $ R^{(0)}_{\phi0}(\varepsilon) $ in the case of magnetopolaronic MRL-model at $ \phi\neq 0$ reaches its maximal value (being equal to $ \exp(-4\phi^{2}) $) at nonzero energy value (which is approximately equal to $ \Gamma_{0} $, in the case: $ \phi^{2} \geq 1$). At $ \varepsilon=0 $ (i.e. at Fermi energy) the transmission coefficient $ R^{(0)}_{\phi0}(0) $ is equal to zero, in contrary with the situation without any quantum vibrations ($ \phi=0$), while at: $ \varepsilon \gg \Gamma_{0} $ it decreases smoothly, similarly to the case $ \phi=0$. This fact means, that in the magnetopolaronic MRL-system at zero temperature, in the limit: $ \Gamma_{0}\gg \hbar\omega_{0} \rightarrow 0 $, it is impossible to "compensate" the magnetopolaronic blockade of resonant tunneling \cite{OUR1} even by means of very high bias voltage. 
\begin{figure}
\includegraphics[height=10 cm,width=9 cm]{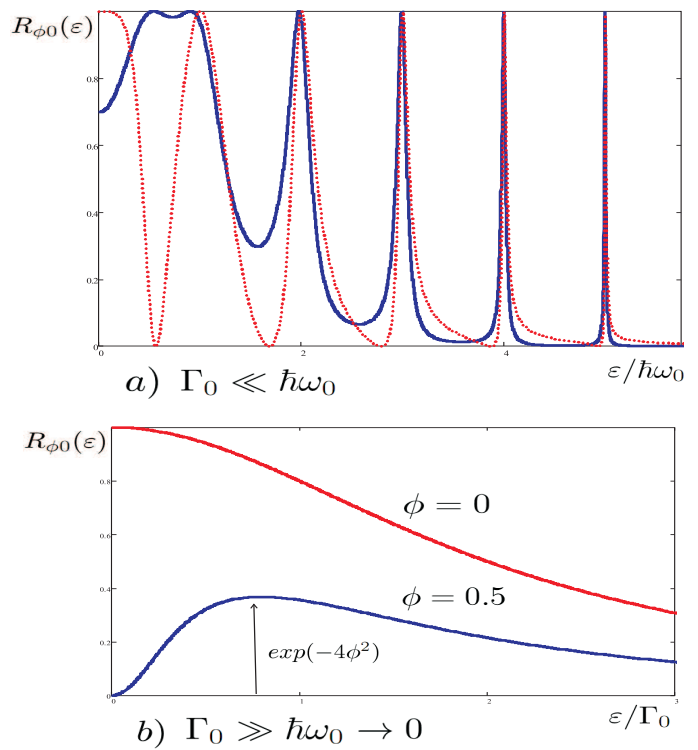}
\caption{a) Low-temperature transmission coefficient (15) as the function of energy (in the units of $ \hbar\omega_{0} $) in the limit: $ \Gamma_{0}\ll \hbar\omega_{0} $ (and $\hbar\omega_{0} \gg T  $) at $ \phi^{2}=\sqrt{3/2} $ (-blue solid line) in comparison with the effective transmission coefficient of corresponding polaronic MRL-model from Ref.\cite{MY} with the same values of all parameters (-red dotted line). Here: $ \Gamma_{0}=0.1\hbar\omega_{0} $; $ T=\hbar\omega_{0} $. 
b) Zero-temperature transmission coefficient (18) in the limit: $ \Gamma_{0}\gg \hbar\omega_{0} \rightarrow 0$, as the function of energy (in the units of $ \Gamma_{0} $) for two cases: i) $ \phi=0$ (-red solid line); ii)$ \phi\neq 0$ (-blue solid line).}
\end{figure}
Obviously, the decoupling of Eqs.(11-13) means, that in the magnetopolaronic MRLM case vibronic subsystem has a complete set of eigenstates, which are the quantum states of a free quantum harmonic oscillator. Thus, corresponding eigenfunctions (i.e. different eigenfunctions of a magnetopolaron) have definite parity with respect to symmetry transformations (5). This parity is equal to $ (-1)^{l} $, where $ \vert l \vert $ is a number of vibrons in $l$-th eigenstate of quantum oscillator. From symmetry operations (5), one can note, that all odd eigenfunctions (corresponding to eigenstates with odd vibron number) should change their sign in the resonant tunneling process, while the sign of all even eigenfunctions remains unchanged.  This could be explained as the consequence of fixed values of the relative Aharonov-Bohm (A-B) phase $ \Delta\varphi^{odd/even}_{l} $ in each $ l $-th vibronic channel of tunneling. This relative A-B phase in $ l $-th channel is equal to $ \pi\vert l \vert $. Thus, resonant "Andreev-like" magnetopolaronic tunneling in the MRL-model turns out to be driven by the relative Aharonov-Bohm phase of value $ \pi $. One could suppose, that in the magnetopolaronic MRL-model the relative A-B phase in each $ l $-th channel of tunneling plays the role of Berry phase \cite{DX,GEF}, which is acquired by real electron in that virtual channel during the process of resonant tunneling \cite{TOP}. 

In conclusion, the exact solution for the average current in magnetopolaronic model of single-electron transistor (SET) with arbitrary magnetopolaronic coupling in vibrating quantum dot is obtained for the first time, in the case of MRLM-realization of magnetopolaronic SET model. Especially, this solution reveal the nontrivial interference between different virtual intermediate vibronic states of magnetopolaron. As the result, at zero temperature, in the magnetopolaronic MRL-model with strong magnetopolaronic coupling, it is impossible to compensate the effect of strong magnetopolaronic blockade by means of bias voltage in the case, where vibron energy is the smallest (nonzero) energy scale in the system. In principle, the qualitatively new effects, being predicted theoretically in the above, could be measured experimentally. One could use these effects for detecting the MRL-type of tunnel coupling in SETs as well as for the estimations of self-frequencies and zero-point-oscillation amplitudes in magnetopolaronic MRL-systems.   

The author thanks to A.Komnik, I.V.Krive, S.I.Kulinich, R.I.Shekhter and F.Pistolesi for valuable discussions.

\end{normalsize}


\begin{references}
\bibitem{VR}All energies in the system are counted from the Fermi energy of TLL leads; in our model it is convinient to put Fermi energy to be equal to zero. Besides that, the "magnetopolaronic" shift of resonant level due to magnetopolaronic coupling is already taken into account in the parameter $\Delta(V_{g})$ \cite{MY,OUR1}.
\bibitem{CUR}It is important to stress out, that although the form of the operator in the right-hand side of Eq.(14) is very similar to that of previously defined current operator (6), operators $ \hat{I}(\infty) $ and  $\hat{I}(0)$ -are differently time-dependent. One could only claim, that in the steady-state: $\langle \hat{I}(\infty)\rangle=\langle \hat{I}(0)\rangle$ due to continuity of average charge density along the system, but the latter equality is valid only on the level of averages, whereas, for corresponding operators it does not take place. Indeed, resonant tunneling of electron in MRLM is always spatially nonlocal and, thus, it results in quantum fluctuations of current only far from the scatterer (i.e. far from the QD).
\bibitem{PTA} It is worth to note, that the effects of tunnel barriers' asymmetry in the magnetopolaronic $ g=1/2 $-TLLRL case are quite the same as ones for corresponding polaronic generalization of  $ g=1/2 $-TLLRL model \cite{MY} and all corresponding formulas could be  obtained, in principle, in the spirit of Ref. \cite{MY} under the assumption about the fermion-boson factorization of averages. But it is important, that for asymmetric magnetopolaronic $ g=1/2 $-TLLRL case the validity of "polaron tunnelling approximation" needs further justification as compared with corresponding polaronic $ g=1/2 $-TLLRL model of Ref.\cite{MY}.
\end{references}
\end{document}